\documentstyle [aps,prl,epsf,multicol]{revtex}

\begin{document}
\draft
\title{Queuing transitions in the asymmetric simple exclusion process}
\author{Meesoon Ha,$^{1}$ Jussi Timonen,$^{2}$ and Marcel {den Nijs}$^{1}$}
\address{$^{1}$Department of Physics, University of Washington,
P.O. Box 351560, Seattle, Washington 98195-1560, USA}
\address{$^{2}$Department of Physics, University of Jyv\"askyl\"a, 
P.O. Box 35, FIN-40351 Jyv\"askyl\"a, Finland}
\date{\today}

\maketitle
\begin{abstract}
Stochastic driven flow along a channel can be modeled 
by the asymmetric simple exclusion process. We confirm numerically 
the presence of a dynamic queuing phase transition at a nonzero
obstruction strength, and establish its scaling properties. 
Below the transition, the traffic jam is macroscopic in the sense that
the length of the queue scales linearly with system size.
Above the transition, only a power-law shaped queue remains. 
Its density profile scales as $\delta \rho\sim x^{-\nu}$ with 
$\nu=\frac{1}{3}$, and $x$ is the distance from the obstacle. 
We construct a heuristic argument, indicating that the exponent
$\nu=\frac{1}{3}$ is universal and independent of the dynamic exponent 
of the underlying dynamic process. Fast bonds create only 
power-law shaped depletion queues, and with an exponent 
that could be equal to $\nu=\frac{2}{3}$, 
but the numerical results yield consistently somewhat smaller values
$\nu\simeq 0.63(3)$. The implications of these results to 
faceting of growing interfaces and localization of directed polymers 
in random media, both in the presence of a columnar defect 
are pointed out as well.

\end{abstract}

\begin{multicols}{2}
\narrowtext

\section{Introduction}
\label{intro}

Queuing is a common nonequilibrium phenomenon in nature. 
It appears, e.g., in stochastic-type driven transport 
through narrow channels, where applications can range from 
electron transport along nanowires to traffic flow on highways. 
Queuing has received a lot of attention from the theoretical side 
for over a decade~\cite{S-Zia,Schuetz}. It is well established that 
many driven flow processes belong to the same universality class as
Kardar-Parisi-Zhang-(KPZ) type growth of 
one-dimensional (1D) interfaces~\cite{K-P-Zhang,KPZgrowth}. In particular, 
the so-called asymmetric simple exclusion process (ASEP)~\cite{Dhar_G-Spohn} 
maps exactly onto the so-called 
body-centered solid-on-solid lattice version of KPZ growth~\cite{BCSOS}. 
This model has been used to describe 
biopolymerization~\cite{MacDonald68_Schuetz97}, 
gel electronics~\cite{Widom91/94}, directed polymers 
in random media~\cite{K-Tang}, traffic jams~\cite{traffic,ourPGM}, 
and the fluctuations 
of shock fronts~\cite{J-Lebowitz92,Schuetz93,D-J-L-Speer,Mallick}.
In the case of periodic boundary conditions,
the time development of the ASEP is exactly soluble 
by the Bethe ansatz~\cite{Dhar_G-Spohn}. For a wider class of setups,  
the exact stationary state has been constructed 
as well using the so-called 
matrix method~\cite{Schuetz,Schuetz93,Mallick,Derrida,S-Domany}.

The starting point and motivation for the study presented here was
actually not queuing in driven flow but faceting in KPZ growth.
Slow flameless combustion of paper produces 1D burning fronts that
evolve in time according to KPZ-type growth. 
The results of early experiments~\cite{early-burn} seemed to deviate 
from KPZ behavior, but more recent investigations demonstrated that 
for length scales larger than about 5-10~mm depending 
on paper structure, random pinning effects do not play a role 
any more, and that the interface obeys 1D KPZ scaling~\cite{recent-burn}.
In these experiments, the paper is impregnated with KNO$_3$ 
to provide the oxygen source necessary 
for maintaining the slow-combustion process. 
The burning speed can also be controlled by the KNO$_3$ concentration.
In particular, the rate can be enhanced or reduced in a narrow strip 
along the burning direction. The latter experiments nicely illustrate 
the presence of nonlinear terms in the equation of motion; 
the burning front facets for enhanced concentrations
but not for reduced ones. The details of these experiments, 
as well as the matching of the experimental data 
with our numerical results for the slow- and fast-bond ASEP 
will be published separately~\cite{Timonen}. 
Here we only present the ASEP queuing perspective.

One of the fundamental questions in driven flow is 
whether a static obstruction, such as a slow bond, 
always results in a traffic jam, 
or whether stochastic fluctuations destroy the queue 
at weak obstacle strengths. Such a vanishing of
the queue as a function of the slow-bond strength represents 
a dynamic phase transition. The size of the queue is 
the order parameter, i.e., it being finite or infinite in length; 
or more precisely, whether the number of
``cars" in the queue scales and diverges with the system size or
remains finite (in obvious analogy with macroscopic occupation of
the ground state in equilibrium Bose condensation).
The existence of such a transition, its scaling properties,
the shape of the density profile near the obstruction,
and also whether information percolates through the slow bond,
are the most important issues.

The ASEP is one of the simplest nonequilibrium driven dynamic processes 
displaying queuing phenomena. Particles move stochastically 
along a chain of sites $1\leq x\leq N_s$, only in one direction, 
with hopping probability $p$ under the constraint that 
the particles can neither pass each other nor occupy the same site, 
$n_x=0,1$. We use random sequential updating of the sites.
The obstacle is introduced by modifying the hopping probability to 
$rp$ at one specific bond along the chain.
$0\leq r<1$ represents a slow bond and $r>1$ a fast bond.
We choose open boundary conditions with the special bond 
in the middle of the chain.

%%%% changes July 14 from here
Mean-field theory predicts an infinitely long traffic jam 
for all $r<1$ and only a logarithmic depletion density profile 
near a fast bond~\cite{W-Tang}. The literature is confused 
about what happens in reality. The notion of an $r_c<1$ queuing transition
seems to have been implicitly presumed in the  ASEP literature 
for over a decade. Kandel and Mukamel~\cite{K-Mukamel} presented early
numerical data (for a different but related growth model
involving parallel updating and  polynuclear growth) suggesting 
a critical point which would correspond to 
$\approx r_c\simeq 0.7$ in our model. Above the queuing transition, 
they reported evidence of continuously varying exponents 
in the density profiles. Some of these aspects have been confirmed 
by more recent studies~\cite{Slanina,J-Lebowitz94}, 
but to the best of our knowledge, 
the existence of an $r_c<1$ has not been resolved unambiguously.
For example, in the ASEP studies with a slow bond and 
periodic boundary conditions, Janowsky and Lebowitz presented 
their phase diagrams as if $r_c=1$ 
(the mean-field location)~\cite{J-Lebowitz92}.
In retrospect $r_c\simeq 0.8$ seems to us consistent with,
e.g., their series expansions~\cite{J-Lebowitz94}.
Their main focus was however elsewhere, 
with the fluctuations in the position of the shock front of the queue, 
at the far end from the slow bond, and not with the precise value of $r_c$.

The exact form of the stationary state in the ASEP can often 
be obtained from the so-called matrix 
and Bethe ansatz method~\cite{Schuetz,Schuetz93,Mallick,Derrida,S-Domany},
but these analytic techniques typically work only 
for very specific boundary conditions and/or update rules.
Sch{\"u}tz~\cite{Schuetz93,Schuetz-extra}, for example, found $r_c=1$
for periodic boundary conditions and parallel updating.
That is consistent with our results, 
because parallel updating creates intrinsically 
weaker stochastic noise than random sequential updating~\cite{all-updates}.
%%%% until here

The directed polymer community was focused on the slow-bond issue 
in the mid 1990s~\cite{Tang,Balents,Kinzelbach,Hwa,Straley,Lassig}.
The driving force behind these studies was 
the realization of such directed polymers 
in terms of flux tubes in type-II dirty superconductors.
The (1+1)-dimensional ASEP is equivalent to 
(1+1)-dimensional KPZ-type growth, and the latter to
a directed polymer in two dimensions subject to a random potential.
In these equivalences, the slope of the KPZ interface is 
the deviation of the local density from a half filling in the ASEP, 
%%%%%%change 1%%%%%%%%%%%
$\partial h/\partial x= 1-2\rho(x)$,
%%%%%%%%%%%%%%%%%%%%%%%%%% 
and the mapping of the KPZ equation to the directed polymer problem
involves the celebrated Hopf-Cole transformation, 
$W=\exp[(\lambda/2\nu) h]$ (with $\lambda$ and $\nu$ the
KPZ coupling constants; for a review see Ref.~\cite{Lassig}).
The slow bond transforms into a columnar defect 
with a short-ranged attractive interaction,
and the queuing issue translates into 
whether the polymer becomes localized to it immediately 
or only beyond a critical defect strength.
The latter is true above a critical dimension $D_c$.
Power counting in the KPZ equation 
and associated field-theoretical renormalization studies suggest that 
$D_c=1$, i.e., our ASEP model is at the critical dimension.
In such cases one expects that $r_c=1$, likely accompanied by 
essential singularities~\cite{Tang,Balents,Kinzelbach,Hwa,Straley,Lassig}.
Our results presented below seem to contradict 
these field-theoretical studies, 
but actually only do so in a limited sense. 
We find a more complex structure.
The queued ASEP phase represents the strongly localized state.
It exists only beyond a critical defect strength $r_c<1$.
The power-law shaped profile that remains for weaker slow bonds, 
represents a form of weak localization. Earlier numerical studies 
in the directed polymer representation confirmed localization 
in $D=D_c=1$ for all $r<1$~\cite{Tang,Balents} 
but were likely insensitive to this distinction.

Faced with the realization of this process 
in terms of slow combustion of paper, our first goal is 
to settle the location of $r_c$ for the sequential update rule,
numerically as accurately as possible. We demonstrate here 
the presence of a dynamic phase transition at $r_c\simeq 0.80(2)$. 
The queue remains infinite in length all the way up to $r_c$.
Its density $\rho_b=\frac{1}{2}(1+\Delta_b)$ decreases as 
$\Delta_b\sim |r_c-r|^\beta$  with $\beta\simeq 1.46(4)$. 
This is presented in Sec.~\ref{rc} after a detailed discussion
of our choice of boundary conditions in Sec.~\ref{bc}.

Having settled the existence of the critical point,
we turn, in Sec.~\ref{dp}, our attention to the density profile 
near the slow bond, $\rho(\tilde x)=\frac{1}{2}[1+\Delta(\tilde x)]$, 
with $\tilde{x}$ the distance from the special bond.
It follows always a power law, 
$\Delta(\tilde{x})\simeq \Delta_b +A\tilde{x}^{-\nu}$.
Below the transition, $r<r_c$, 
the density profile has a power-law tail 
with exponent $\nu=\frac{1}{2}$. 
Above the transition, $r_c<r<1$ and $\Delta_b=0$,  
the density profile has a power-law shape with exponent $\nu=\frac{1}{3}$. 
Notice that the remaining power-law 
queue above $r_c$ is still
infinite in magnitude,
since $\int \Delta (\tilde x) d\tilde x\sim N_s^{2/3}$ diverges.
In the fast-bond scenario, the queue has always a power-law profile
(i.e., $\Delta_b=0$) for all values $r>1$,
with exponent $\nu\simeq 0.63(3)$.

These power-law density profiles are very intriguing,
in particular, the fast-bond one.
They are different from the density profiles near reservoirs,
e.g., those in the exact solution of Derrida {\it et al.}~\cite{Derrida} 
for the open boundary conditions with two reservoirs.
Those have exponential tails in the reservoir-dominated phases, 
and power-law tails with exponent $\frac{1}{2}$ 
in the bulk-dominated maximal-current phase. 
It is well known how to explain these reservoir-related profiles 
with the help of simple scaling arguments involving 
the dynamic exponent $z=\frac{3}{2}$, 
the roughness exponent $\chi=\frac{1}{2}$, 
and the absence or presence of a nonzero group velocity for fluctuations
(see, e.g., Ref.~\cite{ourPGM}).

In Sec.~\ref{nu}, we generalize these heuristic arguments
to the density profiles near the slow bond.
This reproduces the observed slow-bond values
$\nu=\frac{1}{2}$ and  $\nu=\frac{1}{3}$.
A similar line of reasoning for the fast-bond profile
yields $\nu=1/z=\frac{2}{3}$, but is on shaky grounds,
in particular, in the light of the fact that our
numerical results give systematically a somewhat smaller value.

Another important result of this study, presented in Sec.~\ref{ucp},
is that the passage of particles through the slow bond sets itself up 
in the stationary state as an uncorrelated process 
both below and above $r_c$. Fluctuations travel away from it from both sides.
No information passes through the slow bond. The fast bond, on the other 
hand, acts very much like a normal site, and fluctuations flow through it.
In Sec.~\ref{summary}, we summarize our results.

\section{boundary conditions}
\label{bc}

The choice of boundary conditions is important 
for the accuracy of our numerical analysis. The special bond
creates a power-law shaped density profile. Therefore we need to fully
control other sources for density profiles and minimize interference.
We have chosen open boundary conditions with the special bond 
halfway along the road, and particle reservoirs on either side, 
at $x=1$ and $x=N_s$. Particles can only hop to the right, $x \to x+1$. 
The hopping probability is equal to $p$ for all sites,
except for three special sites. The probability to hop 
through the special bond is equal to $p^\prime=rp$.
The probability to enter (leave) the road from (into) the reservoir 
at site $x=1$ ($x=N_s$) is equal to $\alpha p$ ($\beta p$).
%%%%change 2 %%%%%%%%%%%%%%
It is advantageous to set $p=1$, if possible, 
to maximize the speed of the Monte Carlo (MC) simulations, 
but in our case that would exclude us from addressing 
the fast-bond scenario $r>1$. We set $p=\frac{1}{2}$ throughout this study.
There are several alternatives that can speed up the simulations, 
but we did not feel the need to explore them in this study. 
For an example, one might set $p=1$ everywhere along the chain
except at the special bond by increasing 
the update probability of that bond.
%%%%%%%%%%%%%%%%%%%%%%%%%%%%% 

Our choice to employ open boundary conditions might appear surprising.
They often introduce edge effects (surface critical phenomena)
that are typically more difficult to interpret and control 
than those for periodic boundary conditions.
In the ASEP, however, periodic boundary conditions introduce 
a shock wave in the density profile at halfway around the chain, 
opposite to the slow bond. Janowsky and Lebowitz~\cite{J-Lebowitz92} 
studied the fluctuations in the position of this shock wave 
and found for it to fluctuate critically, 
implying the absence of a characteristic length scale. 
We like to decouple the slow bond from those fluctuations
and do so by using open boundary conditions.
The trade-off are density profiles near the edges of the road
induced by the particle reservoirs. But these are under full control 
with the help of the exact solution of the $r=1$ ASEP 
with open boundary conditions~\cite{Derrida}.
The density profiles near the edges have only exponential tails,
provided that we choose suitable values for $\alpha$ and $\beta$. 
At $r=1$, for $\alpha=\beta=\frac{1}{2}$, 
the density profile is completely flat and featureless; 
$\rho_x=\frac{1}{2}$ for all $x$~\cite{Derrida}. 
Thus, we select $\alpha=\beta=\frac{1}{2}$ 
throughout this study~\cite{boundary-free}

Assume that the slow bond creates an infinite queue. 
In the bulk of that queue, far from both the slow bond and the road edge, 
the stationary state is uncorrelated, because locally it is indistinguishable
from a setup with periodic boundary conditions without the slow bond.
Moreover, from the perspective of the sites near the $x=1$ edge, 
this situation is indistinguishable from a setup with $r=1$ (no slow bond) 
where an exit probability $\beta$ at the opposite site of the road could 
be responsible for this enhanced bulk density $\Delta_b$.
From the exact solution of that setup, 
we know that the density profile near the entry edge is exponential,
$\rho(x)\simeq \frac{1}{2}(1+\Delta_b)+B\exp(-x/\xi)$
with a finite correlation length $\xi\sim \Delta_b^2$.
Our simulations confirm this~\cite{Kolomeisky}.

Other important features of the density profile are predetermined as well.
For all $\alpha=\beta$, the density profile has particle-hole symmetry 
with respect to the special bond, $\rho(x)= 1-\rho(N_s+1-x)$. 
Define $\Delta(x)$ as the deviation of the density 
from $\frac{1}{2}$, $\rho(x)=\frac{1}{2} [1+\Delta(x)]$, 
such that $\Delta(x)=-\Delta(N_s+1-x)$. Then, $\Delta_b$ is
the order parameter of our model and represents 
the spontaneous faceting angle of the slow-combustion interface profile 
in the KPZ interpretation.

In the steady-state limit, the current along the chain must be uniform.
From the fact that the bulk stationary state is uncorrelated,
it follows immediately 
that its value through such a bulk bond is equal to
\begin{equation}
\label{current}
J= p \langle \hat n_x(1-\hat n_{x+1})\rangle
=\frac{p}{4}(1-\Delta_b^2).
\end{equation}
This means that we have the option to determine $\Delta_b$
by measuring $J$. The current from the reservoir to the first site
\begin{equation}
\label{edgecurrent}
J = \alpha p \langle(1-\hat n_{1})\rangle
=\frac{\alpha p}{2}(1-\Delta_1)
\end{equation}
must be equal to the bulk current in the stationary state.
This yields, for $\alpha=\frac{1}{2}$,
that the density at site $x=1$ is equal to $\Delta_1=\Delta_b^2$.

The density profiles in front and beyond the special bond
obey particle-hole symmetry, $\langle n_L\rangle =1- \langle n_R\rangle$,
with $x_L$ and $x_R$ the sites immediately 
in front and beyond the special bond.
Moreover, we will demonstrate in Sec.~\ref{ucp} that the ratio 
${\cal R}=\langle \hat n_L \hat n_R\rangle/\langle \hat n_R\rangle $ 
is very close to ${\cal R}=\frac{1}{2}$ for all values of $r$. 
The current through the special bond
\begin{equation}
\label{SBcurrent}
J= rp\langle \hat n_L (1-\hat n_R)\rangle
= rp\left[(1+{\cal R})\langle n_L\rangle -{\cal R}\right]
\end{equation}
must again be equal to the bulk current, 
thus anchoring the value of the density immediately 
in front of the special bond to the order parameter of our process 
as $\Delta_L= \frac{1}{3}[(1-\Delta_b^2)/r-1]$ 
if ${\cal R}$ is exactly equal to $\frac{1}{2}$.

\begin{figure}
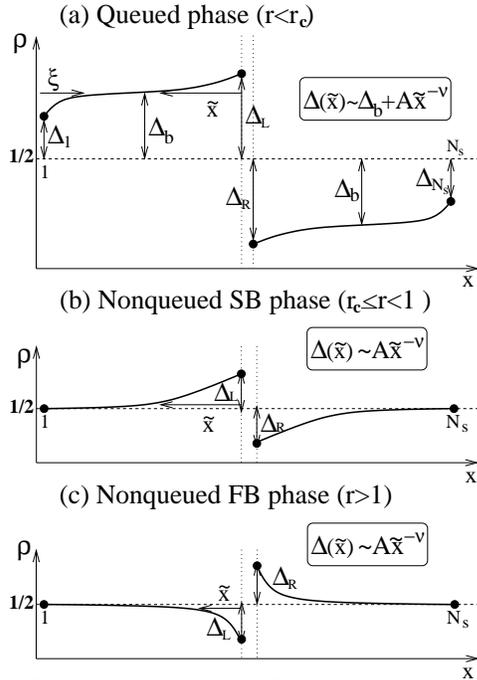

\centerline{\epsfxsize=6.25cm \epsfbox{fig1a.epsi}}
\centerline{\epsfxsize=6.25cm \epsfbox{fig1b.epsi}}
\centerline{\epsfxsize=6.25cm \epsfbox{fig1c.epsi}}
\caption {Schematic density profiles are shown: for the slow bond (SB)
(a) $r<r_c$ (queued phase) and (b) $r_c\le r<1$ (nonqueued SB phase)
and for the fast bond (FB) (c) $r>1$ (nonqueued FB phase).}
\label{profile-schematic}
\end{figure}

Figure~\ref{profile-schematic} summarizes the above discussion.
The current, the densities at the first and last sites, the characteristic
exponential length scale of the density profile near the reservoir edge
(in the faceted phase), and the bulk density are all linked to each other.
This leaves only the spontaneous creation of a nonzero $\Delta_b$ and
the density profile near the special bond as independent issues.

\section{queuing phase transition}
\label{rc}

The first issue at hand is to settle 
by numerical means the burning question 
whether there exists a queuing transition at an $r_c<1$.
We perform MC simulations for system sizes up to $N_s=4096$ 
and analyze them by finite size scaling (FSS) techniques.
The order parameter of the queuing transition is the offset of
the density in the bulk, $\Delta_b$ 
(far from both the special bond and the edge). 
There are several ways to measure this.

\begin{figure}
\centerline{\epsfxsize=6.5cm \epsfbox{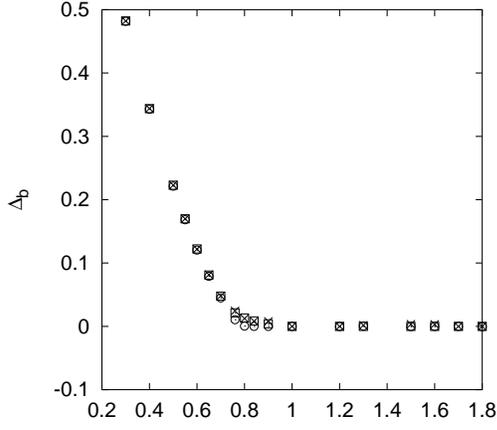}}
\caption{The order parameter $\Delta_b$ vs the strength
$r$ of the special bond at $N_s=4096$, determined from three different
datasets: the average current $J$ (squares),
the density $\Delta_1$ at the first site near the reservoir edge (crosses),
and three parameter power-law  fits to the density profiles near the
special bond (circles).}
\label{orderparameter}
\end{figure}

One can directly measure the density at a site such as $x=\frac{1}{4}N_s$ 
and perform a FSS analysis to determine the asymptotic value. This works, 
but neither $x=\frac{1}{4}N_s$ nor any other fixed site is optimal for
such an analysis, because the density profile has a power-law tail 
at the special bond side and only an exponential one near the reservoir. 
Instead, we show in Fig.~\ref{orderparameter}
the values for $\Delta_b$ from a power-law density profile fit
\begin{equation}
\label{profile}
\Delta(\tilde{x})\simeq  \Delta_b + A \tilde{x}^{-\nu}
\end{equation}
(with $\tilde x=x_R-x$ the distance from the special bond)
at our maximum system size $N_s=4096$.

As pointed out in the preceding section, 
$\Delta_b$ is directly linked to various other quantities, 
such as the average current $J$, the density at the first site 
near the edge $\Delta_1$, and the characteristic length $\xi$ 
of the exponential tail in the density profile near the edge.
We measure these quantities as well, 
and translate the first two into their predictions for $\Delta_b$. 
The results, also shown in Fig.~\ref{orderparameter},
are almost indistinguishable from those 
of the power-law density profile fits.
This confirms our analysis of the preceding section.

\begin{figure}
\centerline {\epsfxsize=6.5cm \epsfbox{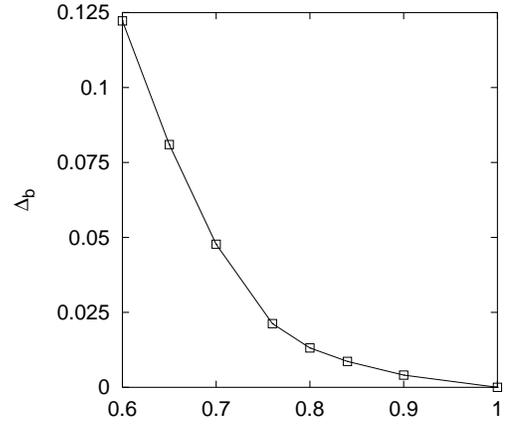}}
\caption{The order parameter $\Delta_b$ vs the strength
$r$ of the slow bond in the vicinity of the
critical point $r_c$, as obtained from the average current dataset.}
\label{order-zoom}
\end{figure}

\begin{figure}
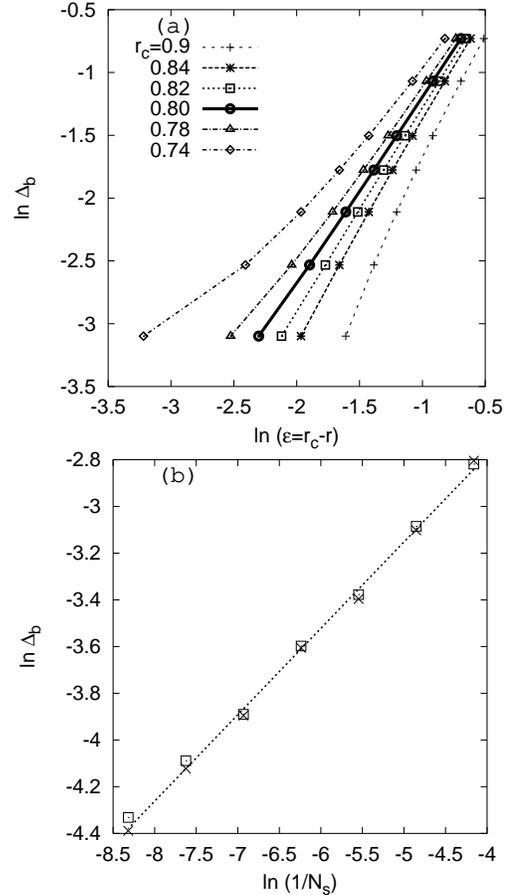

\centerline{\epsfxsize=6.5cm \epsfbox{fig4a.epsi}}
\centerline{\epsfxsize=6.35cm \epsfbox{fig4b.epsi}}
\caption {Determination of the critical point and
critical exponents. (a) Double logarithmic
$\Delta_b\sim|\epsilon|^\beta$ type plots of
the order parameter with $\epsilon= r_c-r$ at $N_s=4096$ for various
choices of $r_c$. The best straight line is found at $r_c=0.80(2)$ and 
with slope $\beta=1.46(4)$. (b) Double logarithmic plots of $\Delta_b\sim 
N_s^{-x_{\Delta}}$ as a function of system size $N_s$ at $r_c=0.80$. The 
slope (dashed line) yields $x_\Delta=0.370(5)$. For clarity we show only 
the data for $\Delta_b$ obtained from $J$ (squares) and $\Delta_1$ 
(crosses).}
\label{exponents-FSS}
\end{figure}

\begin{figure}
\centerline{\epsfxsize=6.5cm \epsfbox{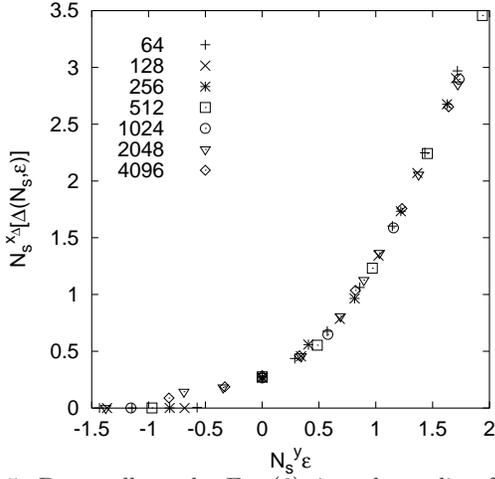}}
\caption{Data collapse by Eq.~(\ref{scaling-function}), i.e., the
scaling function of the order parameter using the values $r_c=0.80$, 
$x_\Delta=0.370$, and $\beta=1.46$
%(i.e., $y=x_\Delta/\beta=0.253$).
as found in Fig.~\ref{exponents-FSS}}
\label{collapse}
\end{figure}

Figure~\ref{orderparameter} suggests very strongly 
the existence of a critical point at about $r_c\simeq 0.8$. 
However, this is a common optical illusion, 
which vanishes upon zooming-in to this point, 
as in Fig.~\ref{order-zoom}. 
Assume that the order parameter obeys the conventional FSS scaling form
\begin{equation}
\label{order-scaling}
\Delta_b(N_s,\epsilon) = b^{-x_{\Delta}}\Delta_b (b^{-1}N_s, b^y \epsilon),
\end{equation}
where $\epsilon=r_c-r$. We test how well our numerical data
obey this scaling relation and what the best values of $r_c$,
$x_{\Delta}$, and $\beta=x_{\Delta}/y$ are. 
The order parameter should scale as a function of $\epsilon$ 
as $\Delta_b\sim\epsilon^\beta$.
In Fig.~\ref{exponents-FSS}, we show a log-log plot of $\Delta_b$
versus $\epsilon$ for various choices of $r_c$ at $N=4096$. 
The best straight line is obtained for $r_c=0.80(2)$ 
with slope $\beta=1.46(4)$. At the same choice for $r_c$, 
the order parameter also scales perfectly as a power law 
$\Delta_b\sim N_s^{-x_{\Delta}}$ 
with a critical dimension $x_\Delta=0.370(5)$.
One should always be on guard for corrections to scaling.
For that reason we plot in Fig.~\ref{collapse}
the scaling function ${\cal S}$ defined as
\begin{equation}
\label{scaling-function}
\Delta_b(N_s,\epsilon) = N_s^{-x_\Delta} {\cal S}(N_s^y \epsilon)
\end{equation}
for $r_c=0.80$, $x_\Delta =0.370$, and $\beta=x_\Delta/y=1.46$.
The data collapses very well, implying only minor corrections to scaling.

\begin{figure}
\centerline{\epsfxsize=6.5cm \epsfbox{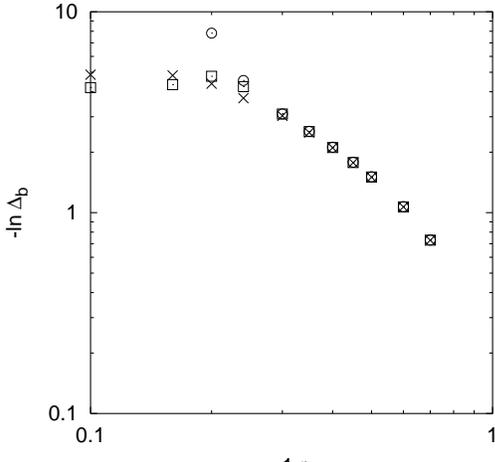}}
\caption{The same data as in Fig.~\ref{orderparameter} 
fitted to a scaling form of the type $\Delta_b$ 
($\Delta_b\equiv\exp[-a(1-r)^b]$) represent a so-called 
essential-singularity characteristic for a possible 
infinite-order-type transition with $r_c=1$. The curves fail to 
straighten out, indicating this is a poor fit.}
\label{singular}
\end{figure}

An alternative scaling form to consider 
is an exponential essential-singularity-type infinite-order transition,
in particular, with $r_c=1$, as suggested by the directed polymer 
renormalization studies~\cite{Tang,Balents,Kinzelbach,Hwa,Straley,Lassig}.
We tried these forms, shown in Fig.~\ref{singular}. They fit our MC data 
poorly.

\section{density profiles}
\label{dp}

\begin{figure}
\centerline{\epsfxsize=6.5cm \epsfbox{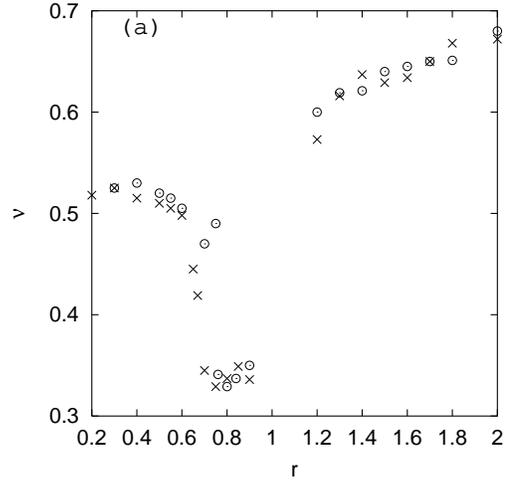}}
\hspace*{-0.6cm}\centerline{\epsfxsize=6.9cm \epsfbox{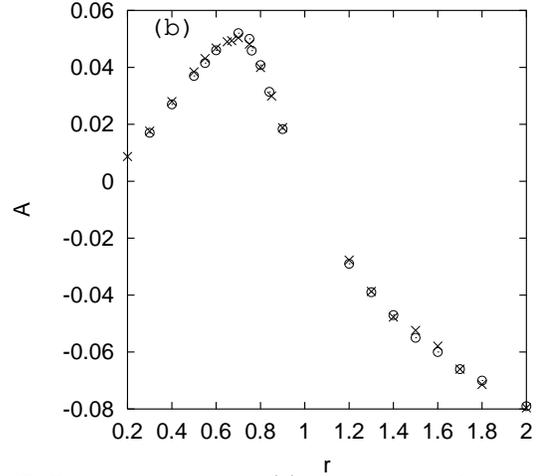}}
\caption{The exponent $\nu$ (a) and the amplitude $A$ (b)
of the power-law shaped density profile, defined in Eq.~(\ref{profile}),
for various $r$ at system sizes $N_s=2048$ (crosses) and $N_s=4096$ (circles).}
\label{nu-A}
\end{figure}

The density profiles near the special bond 
have a power-law shape for all values of $r$. 
Figure~\ref{nu-A} shows our numerical results for the exponent $\nu$ and 
the amplitude $A$ as defined in Eq.~(\ref{profile}). We performed also
two-parameter fits after determining $\Delta_b$ independently
from the numerical values for the current and the density at site $x=1$, 
using the inter-relations outlined in Sec.~\ref{bc}.
These results are identical within the MC noise.

The jumps in $\nu$ at $r_c$ and $r=1$ are very pronounced in Fig.~\ref{nu-A}.
It seems safe to conclude, surely as a starting assumption 
for the discussion in the following two sections, 
that for slow bonds the exponent takes the value $\nu=\frac{1}{2}$ 
in the $r<r_c$ macroscopic queued phase and $\nu=\frac{1}{3}$ 
in the $r_c<r<1$ power-law queued phase.
To the best of our knowledge only one earlier study, 
the one by Slanina and Kotrla~\cite{Slanina}, 
observed this type of power law, but they suggested 
a value different from $\nu=\frac{1}{3}$.
We will present convincing heuristic analytic derivations 
for our values in Sec.\ref{nu}.

The power-law queue for fast bonds, $r>1$, is quite intriguing.
This is where Kandel and Mukamel~\cite{K-Mukamel} sighted 
a possible continuously varying $\nu$. We interpret our data 
as strong evidence for a nonvarying constant value $\nu\simeq 0.63(3)$.
The drop in the estimates in Fig.~\ref{nu-A} near $r=1$ resembles
conventional (multicritical-type) crossover scaling, 
but we cannot verify this explicitly, 
because this power-law decays much faster than 
in both slow-bond phases, and, e.g., at $r=1.1$, 
the amplitude sinks underneath our MC noise level already 
at about $x\simeq 60$. We exclude $x<20$ from our fits 
to avoid (short-distance-type) corrections to scaling.

Obviously we would like to ``talk" this fast-bond power-law profile 
towards $\nu=\frac{2}{3}$, since that number occurs 
naturally in 1D KPZ-type processes. However, our heuristic argument 
for $\nu=1/z=\frac{2}{3}$ is not very strong, see Sec.\ref{nu},
and the fits to the MC data in Fig.~\ref{nu-A} remain consistently 
below that value.

\begin{figure}
\centerline{\epsfxsize=6.5cm \epsfbox{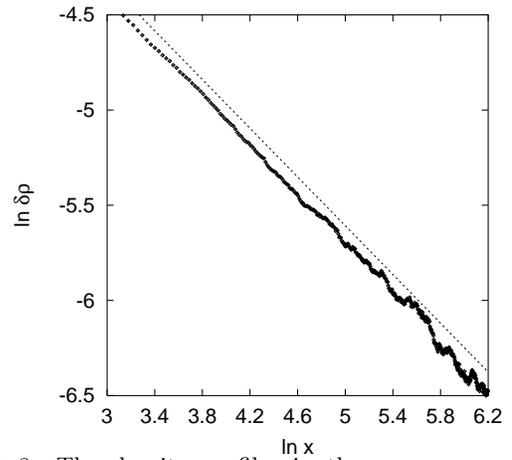}}
\caption {The density profiles in the $r\to \infty$ model at $N_s=4095$,
implemented as a normal chain with uniform hopping probability $p=1$, but
one special double occupancy site in the middle. The dashed line,
with slope $\nu=0.64$, serves as guide to the eye; 
a slope $\nu=2/3$ seems too steep.}
\label{r-infty}
\end{figure}

For this reason we also studied the following $r\to\infty$ like setup. 
Consider a normal chain without any special bond 
but with the site in the middle allowed to be doubly occupied, 
$n_{N_s/2}=0,1,2$. In this setup 
we can increase the hopping probability $p$ to $p=1$ and thus 
speed up MC simulations 
($\alpha$ and $\beta$ are again set equal to $\frac{1}{2}$).
The log-log plot of the density profile, shown in Fig.~\ref{r-infty},
is quite straight. Still, the slope suggests a somewhat smaller exponent,
$\nu=0.64(2)$ (using  $20<x<500$ as the fitting range). 
$\nu=\frac{2}{3}$ is still a possibility, 
but seems to require a significant subdominant correction 
to scaling power-law term. A more detailed analysis becomes 
meaningful only when the noise level is brought down 
by at least one more order of magnitude from our current 
$\delta\rho/\rho\simeq 0.001$ level, which requires vastly longer MC runs.

\section{uncorrelated passage}
\label{ucp}

Our numerical observation that the density profile 
near the special bond follows always a power law, 
$\delta \rho\sim \tilde x^{-\nu}$, in all three phases, 
is far from obvious. The actual values for $\nu$ are even more intriguing. 
In this and the following sections, 
we present intuitive heuristic explanations for the slow-bond values,
and also address the fast-bond case. An important ingredient in this is 
that the passage through the slow bond is an uncorrelated random process
in both the macroscopic queued phase at $r<r_c$ 
and the power-law queued phase at $r_c<r<1$.

In the macroscopic queued phase, 
the absence of passage correlations is easily understood. 
Fluctuations travel away from the slow bond, both in front and beyond it. 
The group velocity of fluctuations $v_g$ points away
from the slow bond in both directions. $v_g=\delta J /\delta\rho$
represents the local response of the current to a density fluctuation.
The stationary state is uncorrelated inside the bulk, 
such that the current is equal to $J=p\rho_b(1-\rho_b)$ 
and $v_g=p(1-2\rho_b)=-p\Delta_b$.
Fluctuation-type wave packets travel with this velocity along the road,
while they broaden spatially as $\xi\sim t^{1/z}$, 
with the 1D KPZ dynamic exponent $z=\frac{3}{2}$. 
In the KPZ growth context, $\Delta_b$ represents
the average slope of the growing surface, 
and the traveling wave packet reflects that the interface moves 
perpendicular to the local surface orientation.

The precise form of $v_g$ for spatially varying densities 
$\rho(x)$ is more complex, but for slowly varying ones, 
like here, we can assume $v_g$ is well represented by
\begin{equation}
\label{group velocity}
v_g(x)=p[1-2\rho(x)]=-p\Delta(x).
\end{equation}
$\Delta(x)$ is positive in the macroscopically queued phase, 
such that the center of mass of a fluctuation packet moves away
from the slow bond linearly in time, $x_{\rm CM}\sim t$. 
During this process, it spreads over a width $\xi\sim t^{1/z}$. 
Fluctuations detach from the slow bond,
because the center of mass of the packet propagates 
faster than its broadening front.
Therefore the density fluctuations at the slow bond 
are uncorrelated in time. No memory remains 
at the slow bond of anything happening there before.
No information passes through the slow bond.

Most of this remains true in the power-law queued phase at $r_c<r<1$.
Now the group velocity vanishes in the bulk, 
but remains nonzero near the slow bond, 
because of the power-law shaped density profile of the queue.
The center of mass of a fluctuation packet still 
moves away from the slow bond, but only as $x_{\rm CM}\sim t^{1/(1+\nu)}$, 
see Eq.~(\ref{group velocity}). During this, it spreads again 
over a width $\xi\sim t^{1/z}$. Therefore, for all $\nu<z-1=\frac{1}{2}$ 
the packet detaches from the slow bond. $\nu=z-1$ is the critical value. 
Numerically we find $\nu\simeq\frac{1}{3}$, see Fig.~\ref{nu-A}(a). 
So the density profile near the slow bond organizes itself 
in a form where the passage fluctuations through the slow bond
are uncorrelated in time and density fluctuations originating 
on the road do not affect it.

It is useful to test this explicitly by numerical simulations,
in particular, in the power-law queued phase. Time correlators, 
such as the current-current autocorrelation function, 
are the preferred tools for this, 
but unfortunately they do not yield much useful information.
The current-current correlator drops in magnitude 
by two orders within 10 MC time steps,
not only near the slow bond, but everywhere along the road as well.
This reflects that, in KPZ growth, $\delta J \sim N_s^{-\sigma}$
scales with a large $\sigma\simeq 2$.

As a second best choice, we focus instead on spatial correlations
between the densities across the slow bond. Consider the ratio
\begin{equation}
\label{SB_correlator}
{\cal R}=\frac{\langle \hat n_L \hat
n_R\rangle}{\langle \hat n_R\rangle},
\end{equation}
with $\hat n_L$ and $\hat n_R$ the density operators
at the sites immediately in front and beyond the special bond.
For reference, the same type of ratio for two nearest-neighbor sites
anywhere along the road, in the bulk, far from edges and slow bonds,
is equal to $\frac{1}{2}$, because the bulk stationary state is uncorrelated, 
with $\langle \hat n_i \hat n_{i+1} \rangle 
=\langle \hat n_i \rangle \langle \hat n_{i+1} \rangle$.
Near the edges, however, and in particular, inside power-law profiles,
the neighbors are correlated, and the ratio moves away from $\frac{1}{2}$.

\begin{figure}
\centerline{\epsfxsize=6.5cm \epsfbox{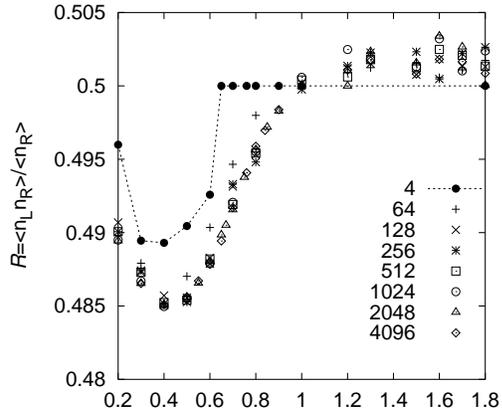}}
\caption {The ratio ${\cal R}$ defined in Eq.~(\ref{SB_correlator}).
${\cal R}$ remains close to the uncorrelated passage value
$\frac{1}{2}$ for all $r$.
The small deviations, of order $3{\%}$, do not scale with system size,
and are mostly described already by a four-sites-type mean-field
approximation, the dashed line.}
\label{SB-corr}
\end{figure}

Figure~\ref{SB-corr} shows that ${\cal R}$, 
as defined in Eq.~(\ref{SB_correlator}), is almost equal to 
$\frac{1}{2}$ for all values of $r$. The deviations are only of order $3\%$. 
This is consistent with the picture that fluctuations travel away 
from the slow bond from both sides.

To quantify this in more detail, 
we consider the following mean-field-type approach, 
in which the road in front and beyond the slow bond are treated as reservoirs 
(devoid of fluctuations as far as the slow bond is concerned).
We solve thus the following two-site problem, 
with only sites $x_L$ and $x_R$ on either side of the slow bond. 
Particles hop onto site $x_L$ with an effective probability 
$\alpha_{\rm eff}p$ from the road in front of it, 
treating the road as a reservoir; 
then move through the slow bond with probability $rp$;
and finally hop away from $x_R$ onto the road beyond the slow bond
with probability $\alpha_{\rm eff}p$, treating that as a reservoir as well.
Finally, we tune $\alpha_{\rm eff}p$ to the value 
where the current takes the same value as in the true system. 
This approximation yields rather trivially ${\cal R}=\frac{1}{2}$ 
for all $r$ and all $\alpha_{\rm eff}$. The dashed line in Fig.~\ref{SB-corr} 
shows ${\cal R}$ for the next level of mean-field theory, with four sites 
instead of two, taking into account local correlations. 
These suffice to reproduce already most of the small deviations 
we observe in the true ${\cal R}$ as a function of $r$, 
and support the uncorrelated passage nature of the process.

The ratio remains equal to ${\cal R}\simeq\frac{1}{2}$ for fast bonds,
and actually even better than for $r<1$. How do fluctuations travel there?
The group velocity changes its sign (moving direction),
because the power-law queue, $\Delta(\tilde x)\simeq A \tilde x^{-\nu}$,
turns into a depletion zone with negative amplitude $A$. 
Fluctuations travel towards the fast bond from both sides. 
It might seem therefore that this passage process must be highly correlated. 
However, fluctuations originating from all over the road 
bombard the fast bond from both sides,
and average each other out. Consider a fluctuation created at a distance $x$ 
from the fast bond. The center of mass of this fluctuation moves towards it, 
and arrives after a time of flight $t\sim x^{\nu+1}$.
During this time, it has broadened over a width $\xi\sim t^{1/z}$.
Ignoring the center of mass movement, 
its leading edge would arrive at the fast bond after a time $t\sim x^{z}$. 
For $\nu>z-1=\frac{1}{2}$ the leading edge arrives well 
before the center of mass, and the latter can be neglected. 
Again, $\nu=z-1=\frac{1}{2}$, is the critical value.
For fast bonds we find numerically $\nu\simeq 0.63$, see Fig.~\ref{nu-A}(a).
This explains why ${\cal R}=\frac{1}{2}$. 
The density profile organizes itself again into a form 
where the passage correlations remain simple.
The fast bond acts very much like an ordinary bulk site, 
and fluctuations flow through it.

\section{a derivation of the density profiles}
\label{nu}

In the preceding section we found that the density profiles
organize themselves into a form such that the passage through
the special bond is an uncorrelated process. 
Here we give heuristic arguments for the actual values of the exponents:
$\nu=\frac{1}{2}$ at $r<r_c$ and $\nu=\frac{1}{3}$ at $r_c<r<1$ for slow bonds.
We also explain why for fast bonds $\nu\simeq 1/z$.

First, consider the slow bond $\nu=\frac{1}{2}$ power-law density profile 
in the macroscopic queued phase at $r<r_c$.
The total number of excess particles in this queue diverges 
with system size $N_s$ as $\delta N\sim N_s^{1/2}$.
This has a familiar ring to it. In the Derrida {\it et al.}~\cite{Derrida} 
type open system setup with reservoirs on both sides and no special bonds,
the fluctuations in the total number of particles on the road 
scale as $\delta N\sim N_s^{1/2}$. In that setup, 
this property does not translate into power-law-type density profiles,
except when the road is half filled, $\rho_b=\frac{1}{2}$.
The density profiles are exponential or featureless 
in the two $\rho_b\neq \frac{1}{2}$ phases
where either reservoir controls the bulk density.

The parking garage process of Ref.~\cite{ourPGM} is closer to queuing 
dynamics. In that study the two reservoirs were merged into one,
such that the road forms a loop 
starting and ending in the same parking garage.
The total number of cars in the system is then conserved,
leading to dynamic phase transitions 
between condensate-type stationary states 
where the garage is macroscopically occupied and
a normal phase where it is not. That process has two parameters,
the total number of cars in the system 
and a modified hopping probability $\alpha p$ to jump 
from the garage onto the first site of the road.

The fluctuations in total number of particles
on the road is again equal to $\delta N\sim N_s^{1/2}$.
The explanation of this goes as follows~\cite{ourPGM}
for the normal phase, the nonmaximal-current condensate phase,
and also at the transition point between them. 
The group velocity of fluctuations $v_g$ is nonzero.
This means that the departure of cars from the garage is 
an uncorrelated process. Fluctuations detach from the garage 
because they travel away faster (linear in time) than 
they are spreading backward (as $\xi\sim t^{1/z}$ with $z=\frac{3}{2}$).
Moreover, after a time of flight $t_{\rm flight} = N_s/v_g$ 
they move around the loop, return to the garage, 
and are completely erased.
So we deal with $t_{\rm flight}$ random uncorrelated deposition events.
The fluctuations in the number of cars on the road therefore scale as
$t_{\rm flight}^{1/2}$. In the condensate phase, 
these fluctuations do not lead to an offset 
in the average density of parked cars, 
because the garage is macroscopically occupied,
and positive and negative fluctuations cancel out against each other.
But at the transition point, the bottom of the garage becomes visible.
This limits the negative density fluctuations, and therefore introduces
a bias towards increased occupation, 
such that the number of parked cars is
enhanced and scales as $\delta N_P\sim N_s^{1/2}$.

The same type of reasoning applies to the slow-bond setup.
The passage through the slow bond is a stochastic uncorrelated event
(as demonstrated in the preceding section), 
similar to departures from the garage mentioned above. 
Again, all memory is erased after $t_{\rm flight}\sim N_s$,
the time that takes for a fluctuation to travel 
from the slow bond to the reservoir.
The fluctuations in the number of cars passing through
therefore scale as $t_{\rm flight}^{1/2}$. These fluctuations are biased again,
because the sites immediately in front and beyond the slow bond 
are not reservoirs. Excess particles waiting to pass are spread out, 
and not available for immediate passage. The passage process is biased, 
because slow bonds process particles slower than normal bonds,
while the passing probability of vacancies does not depend on the value of $r$.
(Our process has particle-hole symmetry but only
in conjunction with left-right mirroring with respect to the special 
bond.)
The total number of excess cars near the slow bond waiting to pass 
scales therefore as $\delta N_P\sim N_s^{1/2}$. 
These extra particles must be accommodated over a stretch of road $x<x_L$ 
behind the slow bond. We can imagine two ways to realize this: 
an exponential density profile with a correlation length diverging 
as $\xi\sim N_s^{1/2}$ or a power-law profile with $\nu=\frac{1}{2}$ 
as we actually observe. The power law is indeed more likely 
given the intrinsic critical nature of ASEP.

Next, let us generalize this argument 
to the $\nu=\frac{1}{3}$ power law above the queuing transition, 
at $r_c<r<1$. The time of flight of a fluctuation to travel 
from the slow bond all the way back to site $x=1$, scales now only as,
$t_{\rm flight}\sim N_s^{\nu+1}$ 
since $v_g= -p\Delta(\tilde{x})\sim \tilde x^{-\nu}$.
Assume that $\nu<\frac{1}{2}$, 
in which case the fluctuations still detach from
the slow bond and the passage through the slow bond remains uncorrelated.
The process at the slow bond is still biased toward low density fluctuations.
FSS corrections to the total number of cars in the queue is proportional to
$t_{\rm flight}$ uncorrelated events:
\begin{equation}
\label{nu=1_3}
\delta N\sim t_{\rm flight}^{1/2}\sim N_s^{(\nu+1)/2}.
\end{equation}
This queue heaps up behind the slow bond, and again arranges
itself in the form of a power-law shaped density profile,
$\delta \rho\sim \tilde x^{-\nu}$. Self-consistency implies that
$(\nu+1)/2=-\nu+1 \to \nu=\frac{1}{3}$, in accordance with the observed value.

The $\nu\simeq 0.63(3)$ power law for the fast bond is more challenging.
This is a fundamentally different phenomenon. 
Again fluctuations travel across the system, but now run towards 
the fast bond instead of away from it. 
Actually, as shown already in the preceding section, for $\nu 
>z-1=\frac{1}{2}$ 
the time of flight of the center of mass of a fluctuation
$t_{\rm flight}\sim N_s^{\nu+1}$ is longer than 
the time it takes that same fluctuation to spread over the entire system 
$t\sim N_s^{z}$. In the $r<1$ phases, we are allowed to ignore 
for this reason the spreading of the fluctuations, 
and only consider their center of mass motion (the time of flight). 
At $r>1$ this is reversed. In the $r<1$ phases, the exponent $\nu$ was 
insensitive to the actual value of the dynamic exponent $z$ of 
the dynamic process. In the $r>1$ phase, it must depend on $z$.

A $\nu\simeq \frac{2}{3}$ power-law tail is very rare. 
It does not appear, e.g., anywhere in the Derrida {\it et al.}~\cite{Derrida} 
type two-reservoirs setup. Interestingly, however, 
this density profile appeared already 
in the parking garage ASEP study~\cite{ourPGM}; 
at the second-type condensation transition, 
from the ``normal" to the ``maximum current" phase.
The characteristic feature was that, at the transition point, 
the garage started to transmit information (seized to act as a reservoir), 
and that at that point the bulk group velocity was zero. 
The similarities with the fast bonds are striking.
We are clearly looking at the same type of phenomenon. 
The slow bond does not transmit information, 
while the fast bond acts very much like a normal site 
and fluctuations move (i.e., they spread) 
through it (see also the preceding section).

What might the true value of $\nu\simeq 0.63(3)$ be?
An obvious guess is that $\nu=1/z$, but how to explain this?
One of the crucial aspects must be again 
that the processing of fluctuations is biased at the fast bond, 
leading to a depletion queue with a total deficit of
$$
\delta N\sim \int_0^{N_s/2} d\tilde x~ \tilde x^{-\nu} \sim
N_s^{1-\nu} \sim N_s^{1/3}
$$
particles. The sign of this in now negative, 
because at the fast bond the particles are processed  
faster than at the normal bonds, 
while the passing rate of vacancies does not depend on $r$.

Density fluctuations are created everywhere along the road,
all the time, and with a common characteristic amplitude. 
Each spreads in time over a region $\xi\sim t^{1/z}$. 
A fluctuation created at a distance $\tilde x$ from the fast bond  
arrives there after a time $t\sim \tilde x^{z}$ 
and with a reduced amplitude (from spreading) of order $A \tilde x^{-1/z}$. 
The asymmetry in processing high and low density fluctuations 
gives rise then to a density deficit of order $A \tilde x^{-1/z}$ 
from fluctuation originating at distance $\tilde x$. 
Next, adopting rather frivolously superposition principle concepts,
one would guess that the total density deficit 
in front of the fast bond scales as 
$\delta N\sim \int_0^{N_s/2} d\tilde x~ A \tilde x^{-1/z}\sim N_s^{1-1/z}$ 
in agreement with what we observed numerically.

Although the last argument is reasonably appealing,
it is certainly not convincing. A more robust explanation is needed. 
Moreover, the numerical value $\nu\simeq 0.63(3)$ is sufficiently lower 
to cast serious doubts that $\nu=\frac{2}{3}$ is correct.
On the other hand, the above argument 
serves as a proper order of magnitude estimate.

\section{summary}
\label{summary}

In this study, we reconfirmed numerically, 
and beyond doubt the presence of a queuing phase transition 
in the ASEP with a slow bond of strength $r_c\simeq 0.80(2)$.
We established the two scaling exponents of this transition.
The order parameter, the excess density in the queue, vanishes as
$\Delta_b\sim |r_c-r|^\beta$ with $\beta=1.46(4)$. At the transition point, 
the number of particles in the queue scales with system size 
as $\Delta_b\sim N_s^{x_\Delta}$ with $x_\Delta=0.370(5)$.

From a more general perspective, the transition illustrates 
that weak obstructions do not give rise to macroscopic traffic jams
(queues with lengths that scale linearly with the system size).
The stochastic fluctuations overwhelm the slow bond above $r_c$.

A second result of our study is that above $r_c$
a power-law shaped queue (traffic jam) remains,
$\delta \rho\simeq A\tilde x^{-\nu}$, with $\tilde x$ 
the distance from the obstruction and opposite signs for $A$ 
in front and beyond the obstruction. 
The exponent is equal to $\nu=\frac{1}{3}$.
This value is most likely universal, 
because our heuristic derivation for the general case 
does not involve specific details of the dynamics.
In particular, it does not involve the KPZ value 
of the dynamic exponent $z$, except for the requirement
that fluctuations travel faster away from the slow bond than they spread. 
The argument applies for any process with  $z>1+\nu=\frac{4}{3}$.

From the directed polymer perspective our results are unexpected.
The slow-bond queuing transition represents
a crossover from strong to 
%%%%change 4%%%%% 
a weaker (but still strong) form of localization,
%%%%%%%%%%%%%%%%% 
because the $\nu=\frac{1}{3}$ power-law density tail 
near the slow bond contains still an infinite number of particles;
(naively) the polymer distribution behaves as $\langle W\rangle 
\sim \exp[(\lambda/2\nu)\langle h\rangle]\sim \exp(-Cx^{1-\nu})$.
It will be interesting to see how this weak localized phase,
and the above exact self-consistent argument for $\nu=\frac{1}{3}$, 
can be integrated and reconciled with the field-theoretical descriptions 
of Refs.~\cite{Tang,Balents,Kinzelbach,Hwa,Straley,Lassig}.

For fast bonds (such as a local widening of the road), 
a macroscopic depletion queue 
(with a length proportional to the road length) never appears. 
Instead, a power-law shaped depletion queue is always present
with exponent $\nu\simeq 0.63(3)$. It remains yet unclear 
whether this value is equal to $\nu=1/z$. 
It will be interesting to study 
how our results extend to other models of 
driven flow along one-dimensional channels, 
in particular, to non-KPZ-type dynamics.

\section*{acknowledgments}

We would like to thank Joachim Krug for helpful discussions.
This research was supported by the National Science Foundation
under Grant No. DMR-9985806, a grant from the Netherlands Organization
for Scientific Research (NWO), and by the Academy of Finland.

%\end{multicols}
%\end{document}

\end{multicols}

\begin{thebibliography}{99}

\bibitem{S-Zia}
B. Schmittmann and R.K.P. Zia,
in {\em Phase Transitions and Critical Phenomena},
edited by C. Domb and J. Lebowitz (Academic Press, New York, 1995),
Vol. 17.

\bibitem{Schuetz}
G.M. Sch{\"u}tz, in {\em Phase Transitions and Critical Phenomena},
edited by C. Domb and J. Lebowitz (Academic Press, New York,
2001), Vol. 19.

\bibitem{K-P-Zhang}
M. Kardar, G. Parisi, and Y.C. Zhang, Phys. Rev. Lett.
{\bf 56}, 889 (1986).

\bibitem{KPZgrowth}
J. Krug and H. Spohn, in {\em Solids far from Equilibrium}, edited
by C. Godr\`{e}che (Cambridge University Press, Cambridge,
England, 1991), and references therein.

\bibitem{Dhar_G-Spohn}
D. Dhar, Phase Transitions {\bf 9}, 51 (1987);
L.-H. Gwa and H. Spohn, Phys. Rev. Lett. {\bf 68}, 725 (1992);
Phys. Rev. A {\bf 46}, 844 (1992).

\bibitem{BCSOS}
P. Meakin, P. Ramanlal, L.M. Sander, and R.C. Ball, Phys. Rev. A
{\bf 34}, 5091 (1986); M. Plischke, Z. R{\'a}cz, and D. Liu, Phys.
Rev. B {\bf 35}, 3485 (1987); J. Neergaard and M. den Nijs, Phys.
Rev. Lett. {\bf 74}, 730 (1995).

\bibitem{MacDonald68_Schuetz97}
J.T. MacDonald, J.H. Gibbs, and A.C. Pipkin, Biopolymer {\bf 6}, 1
(1968); J.T. MacDonald and J.H. Gibbs, {\it ibid}. {\bf 7}, 707
(1969); G.M. Sch{\"u}tz, Int. J. Mod. Phys. B {\bf 11}, 197 (1997).

\bibitem{Widom91/94}
B. Widom, J.L. Viovy, and A.D. Defontanies, J. Phys. I {\bf 1}, 1759
(1991); G.T. Barkema, J.F. Marko, and B. Widom, Phys. Rev. E. {\bf 49},
5303 (1994).

\bibitem{K-Tang}
J. Krug and L.-H. Tang, Phys. Rev. E {\bf 50}, 104 (1994).

\bibitem{traffic}
M. Schreckenberg, A. Schadschneider, K. Nagel, and N. Ito, Phys. Rev. E
{\bf 51}, 2939 (1995).

\bibitem{ourPGM}
M. Ha and M. den Nijs, Phys. Rev. E {\bf 66}, 036118 (2002).

\bibitem{J-Lebowitz92}
S.A. Janowsky and J.L. Lebowitz, Phys. Rev. A {\bf 45}, 618 (1992).

\bibitem{Schuetz93}
G. Sch{\"u}tz, J. Stat. Phys. {\bf 71}, 471 (1993).

\bibitem{D-J-L-Speer}
B. Derrida, S.A. Janowsky, J.L. Lebowitz, and E.R. Speer,
J. Stat. Phys. {\bf 73}, 813 (1993).

\bibitem{Mallick}
K. Mallick, J. Phys. A {\bf 29}, 5375 (1996).

\bibitem{Derrida}
B. Derrida, M.R. Evans, V. Hakim, and V. Pasquier,
J. Phys. A {\bf 26}, 1493 (1993).

\bibitem{S-Domany}
G. Sch{\"u}tz and E. Domany, J. Stat. Phys. {\bf 72}, 277 (1993).

\bibitem{early-burn}
J. Zhang, Y.-C. Zhang, P. Alstr{\o}m, and M.T. Levinsen, Physica A
{\bf 189}, 383 (1992).

\bibitem{recent-burn}
J. Maunuksela {\em et al}., Phys. Rev. Lett. {\bf 79}, 1515 (1997);
M. Myllys {\em et al}., {\it ibid}. {\bf 84}, 1946 (2000);
Phys. Rev. E {\bf 64}, 036101 (2001).

\bibitem{Timonen}
M. Myllys, J. Maunuksela, J. Merikoski, J. Timonen, V.K. Horvath,  M.
Ha, and M. den Nijs, Phys. Rev. E {\bf 68}, 05XXXX (2003) in press
(cond-mat/0307231).

\bibitem{W-Tang}
D.E. Wolf and L.-H. Tang, Phys. Rev. Lett. {\bf 65}, 1591 (1990).

\bibitem{K-Mukamel}
D. Kandel and D. Mukamel, Europhys. Lett. {\bf 20}, 325 (1992).

\bibitem{Slanina}
F. Slanina and M. Kotrla, Physica A 256, 1 (1998).

\bibitem{J-Lebowitz94}
S.A. Janowsky and J.L. Lebowitz, J. Stat. Phys. {\bf 77}, 35 (1994).

\bibitem{Schuetz-extra}
G.M. Sch{\"u}tz, J. Stat. Phys. {\bf 88}, 427 (1997).

\bibitem{all-updates}
For a comparison of the properties of the
ASEP process with various update procedures, see, e.g.,
N. Rajewsky, L. Santen, A. Schadschneidr, and M. Schreckenberg,
J. Stat. Phys. {\bf 92}, 151 (1998).

\bibitem{Tang}
L.-H. Tang and I.F. Lyuksyutov, Phys. Rev. Lett. {\bf 71} 2745 (1993).

\bibitem{Balents}
L. Balents and M. Kardar, Phys. Rev. B {\bf 49}, 13030 (1994).

\bibitem{Kinzelbach}
H. Kinzelbach and M. L®assig, J. Phys. A {\bf 28}, 6535 (1995).

\bibitem{Hwa}
T. Hwa and Th. Nattermann, Phys.Rev. B {\bf 51}, 455 (1995).

\bibitem{Straley}
E.B. Kolomeisky and J.P. Straley, Phys.Rev. B {\bf 51}, 8030 (1995).

\bibitem{Lassig}
M. Lassig, J. Phys. C {\bf 10}, 9905 (1998).

\bibitem{boundary-free}
$\alpha=\beta=\frac{1}{2}$ is the proper choice
to get rid of boundary effects.
For larger values, $\alpha=\beta>\frac{1}{2}$,
like in Ref.~\cite{J-Lebowitz94}, the reservoir edge
density profiles  decay as power laws with exponent $\nu=1/2$,
see Ref.~\cite{Derrida}, such that the boundary effects are strong.

\bibitem{Kolomeisky}
A.B. Kolomeisky, J. Phys. A {\bf 31}, 1153 (1998), recently
studied the same setup as ours for various $\alpha$ and $\beta$.
He limited his study to the exponential density profile near the
reservoir edge, comparing numerical results with mean-field
theory.
\end{thebibliography}
\end{document}